\documentclass[10pt,onecolumn,draftclsnofoot]{IEEEtran}

\usepackage{graphicx}
\usepackage{times}
\usepackage{amsmath}
\usepackage{color}
\usepackage{mathptm}
\usepackage{epsfig}
\usepackage{array}
\usepackage{epsfig}
\usepackage{slashbox}
\usepackage{diagbox}
\usepackage{amsfonts}

\usepackage{url}

\ifCLASSINFOpdf

\else

\fi

\begin{document}
%
\title{Stochastic Optimal Linear Control of Wireless Networked Control Systems with Delays and Packet Losses}

\author{Zhuwei~Wang,
        Xiaodong Wang,~\IEEEmembership{Fellow,~IEEE,}
        and~Lihan~Liu 
\thanks{Z. Wang and X. Wang are with the Electrical Engineering Department, Columbia University,
New York, NY 10027 USA (e-mail: zhuwei@ee.columbia.edu, wangx@ee.columbia.edu).}
\thanks{L. Liu is with the Beijing University of Posts and Telecommunications,
Beijing, 100876 P. R. China (email: lihanliu.nyu@gmail.com).}
}


\maketitle

\begin{abstract}
In this paper, the design of the optimal decentralized state-feedback controllers is considered for a wireless sensor and actuator network (WSAN) with stochastic network-induced delays and packet losses. In particular, taking advantage of multiple controllers, we model the WSAN as a wireless networked control system (NCS) with decentralized controllers, and then formulate the stochastic optimal state-feedback control problem as a non-cooperative linear quadratic (LQ) game. The optimal control law of each controller is obtained that is a function of the current plant state and all past control signals. The performance of the proposed stochastic optimal control algorithm is investigated using both a genetic control system and a load frequency control (LFC) system in power grid.
\end{abstract}


\IEEEpeerreviewmaketitle

\section{Introduction}
\IEEEPARstart{N}{etworked} control systems (NCSs) in which the shared communication medium is used for the connections between the plant and the controller have recently attracted much attention due to their potential applications in various areas such as power grid \cite{IRef1}, dc motors \cite{IRef2}, robotic networks \cite{IRef3}, etc. However, two significant challenges are the network-induced delay and the packet loss, which cause performance degradation and potential system instability. In the literature, network-induced delays have been modeled in various forms such as constant delays \cite{IRef4}, short stochastic delays \cite{IRef5}, and long stochastic delays \cite{IRef6}. The typical approaches to modeling, analysis, and synthesis of NCSs with delays are summarized in \cite{IRef7}\cite{IRef8}. On the other hand, the packet loss can be modeled either as a stochastic process \cite{IRef9} or a deterministic one \cite{IRef4}. If the packet containing the current state information is dropped, the typical solution is to either predict the lost state \cite{IRef13}, or use the previous state values \cite{IRef4}, or simply set the lost state to zero \cite{IRef14}. With full state information, the optimal control problem can be solved. In particular, the stochastic optimal linear quadratic (LQ) controller is developed for the NCS in \cite{IRef5}\cite{IRef6} with short or long network-induced delays. In the presence of both stochastic delays and packet losses, the optimal LQ state-feedback control law is derived in \cite{IRef15}.

Compared to the traditional NCS, wireless NCSs, especially wireless sensor and actuator networks (WSANs), offer architectural flexibility and additional degrees of freedom \cite{IRef16}. Several new standards have recently been introduced for multi-hop WSANs, e.g., WirelessHART \cite{IRef18}, ISA-100 \cite{IRef19}. A significant amount of works have considered WSANs under unreliable wireless communications \cite{IRef20}-\cite{IRef26} and real system applications \cite{IRef27}\cite{IRef28}. In particular, the problem of optimal controller placement in multi-hop WSANs is addressed in \cite{IRef20}\cite{IRef21}, and it is shown that placing the controller at the actuator node achieves better performance than at the sensor node under certain conditions. Ref. \cite{IRef22} exploits the flexibility of control design in WSANs to study the adaptive controller placement. The optimal LQ Gaussian control problem is considered in \cite{IRef23}\cite{IRef24} with the signal estimation over lossy networks.
In \cite{IRef25}\cite{IRef26}, the necessary and sufficient conditions under which the closed-loop WSAN is guaranteed to be stable are studied. Note that, all works mentioned above on both WSAN and wireless NCS focus on the case of the single controller.

As the modern control system becomes more complex and large-scaled, a wireless NCS usually consists of multiple controllers (or players, or agents) to coordinately maintain the stability and improve the performance of the system \cite{IRef29}. Recently, the wireless NCS with decentralized controllers has become an important research topic. Ignoring the network-induced delay, the Pareto optimality and Nash equilibrium solutions for the cooperative and non-cooperative games, respectively, are obtained with a focus on LQ differential games \cite{IRef30}-\cite{IRef32}. With the constant input delay, a sufficient condition on the network-induced delay to guarantee consensus for the decentralized coordination of a multi-controller continuous-time system is presented in \cite{IRef33}.
In \cite{IRef34}, a cooperative medium access control protocol is proposed for the distributed feedback NCS under wireless transmission impairments such as packet delays and losses.
A cross-layer framework for the joint design of wireless networks and decentralized controllers is proposed in \cite{IRef35}, where the centralized control and clock-driven controllers are considered and the total network-induced delay is assumed to be one sampling period.
The stability of a decentralized control strategy is studied in \cite{IRef36}, where the network itself acts as a controller, and each node (including the actuator nodes) performs linear combinations of internal state variables of neighboring nodes. Considering the packet loss, the necessary and sufficient conditions for the stability of the delay-dependent decentralized control system are derived in \cite{IRef37}.
Other works on decentralized control systems with different dynamics and applications have been carried out in \cite{IRef38}-\cite{IRef40}.
Unfortunately, the above works all address stability issues of the NCS with decentralized controllers, but the optimality problem remains unexplored.

Due to the architecture flexibility of WSANs, there is significant potential for the WSAN to take advantage of multiple controllers to cooperatively improve the system performance and stability. In this paper, we address the optimality issue in wireless NCS with decentralized controllers. In particular, the optimal state-feedback control problem is investigated for a linear WSAN with decentralized controllers in the presence of stochastic network-induced delays and packet losses. Using the quadratic cost function, the optimal solution is obtained as a feedback non-cooperative control law, which is linear with the current plant state and all past control signals. The performance of the proposed algorithm is assessed using both a genetic control system and a load frequency control (LFC) system in power grid.

The remainder of this paper is organized as follows. The system model and problem formulation are given in Section \ref{System Model}. We then provide the design of state-feedback controllers with stochastic delays and packet losses in Section \ref{Derivation}. Section \ref{extension} discusses the effect of imperfect information caused by the packet loss. Numerical results and conclusions are given in Section \ref{Simulation} and Section \ref{Conclusions}, respectively.

\section{System Model and Problem Formulation}
\label{System Model}
In this section, we first describe the structure of WSANs, and then cast it as a wireless NCS with decentralized controllers in the presence of stochastic delays and packet losses. Finally, the optimal linear state-feedback control problem is formulated as a non-cooperative LQ game.

\subsection{WSANs with Multiple Controllers}
\begin{figure}
  \begin{center}
  \includegraphics[width=0.65\textwidth]{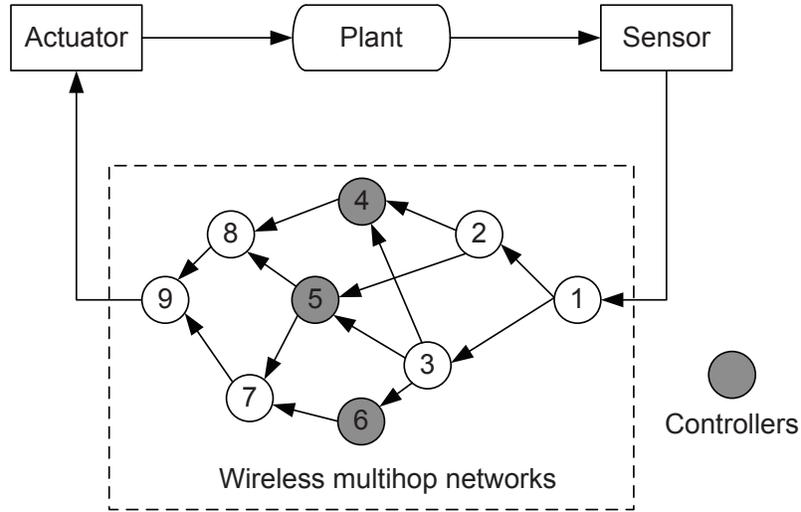}
  \caption{The structure of the WSAN.}
  \label{Fig1}
  \end{center}
\end{figure}

\begin{figure}[h]
  \begin{center}
  \includegraphics[width=0.7\textwidth]{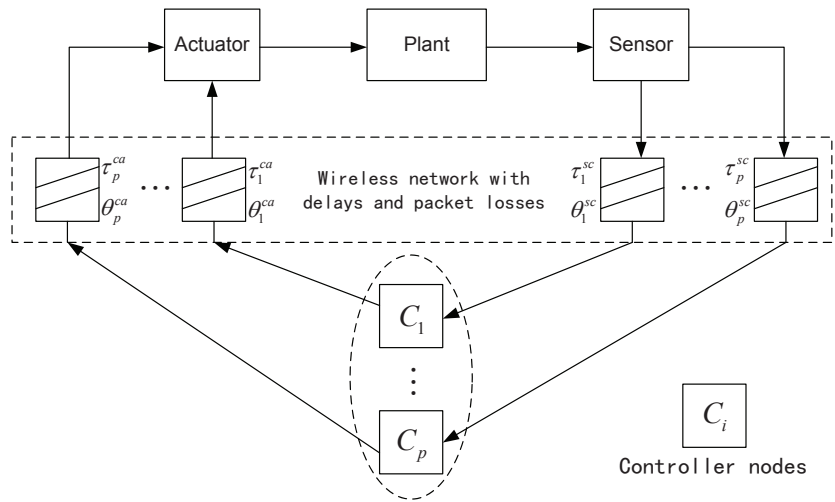}
  \caption{A wireless NCS with multiple ($p$) decentralized controllers.} \label{Fig2}
  \end{center}
\end{figure}

The structure of a WSAN is shown in Fig.~\ref{Fig1}, where the plant, actuator, and a number of sensors (including the sensor and relay nodes) together form a closed-loop NCS. We assume that the plant is a continuous-time linear time-invariant system while all sensors operate in discrete-time. The sampled plant state is sent to the actuator through the wireless multihop network. In the traditional WSAN, only one sensor (or relay) node is selected as the controller to maintain the system stability \cite{IRef20}-\cite{IRef26}. However, with the development of the modern control system, using the cooperative ability of multiple controllers in a decentralized fashion is a potential way to improve the control system performance. As shown in Fig.~\ref{Fig1}, there are three controllers (i.e., node 4, 5, and 6) that coordinately generate control signals which are fed back to the plant. When multiple controllers are considered in the WSAN, Fig.~\ref{Fig1} can be converted to an equivalent wireless NCS with decentralized controllers shown in Fig.~\ref{Fig2},
which consists of five parts: a shared wireless network, the controlled plant, the sensor, actuator, and controllers.

\subsubsection{The Shared Wireless Network}
\label{subsubsection1}
In practice, the transmission in NCSs usually suffers network-induced delays and packet losses. As shown in Fig.~\ref{Fig1} and Fig.~\ref{Fig2}, for a given controller $i$, the sensor-to-controller delay $\tau _i^{sc}$ and the controller-to-actuator delay $\tau _i^{ca}$ should be considered, and $\tau _i  = \tau _i^{sc}  + \tau _i^{ca}$ is the total time delay. The packet loss is modeled by a random switch. In particular, $\theta _i^{sc}$ and $\theta _i^{ca}$ denote the sensor-to-controller and the controller-to-actuator losses, respectively. For example, $\theta _i^{sc}  = 1$ indicates that the sensor packet is successfully transmitted to the $i$-th controller, whereas $\theta _i^{sc}  = 0$ indicates that the packet is lost. Then, $\theta _i  = \theta _i^{sc} \theta _i^{ca}$ denotes the packet loss from the sensor to the actuator through controller $i$.

In this paper, both network-induced delays and packet losses are taken into consideration, and all of them are assumed to be stochastic such that only their distributions are known \cite{IRef5}\cite{IRef6}\cite{IRef15}. In addition, the total time delay $\tau _i$ is assumed to be smaller than one sampling period \cite{IRef5}\cite{IRef40}. This is reasonable because we consider the packet loss, which is one main reason to cause the long delay \cite{IRef7}. Furthermore, the short delay assumption ensures the sensor signal arrives at the controller in the order of their sampling times to avoid the disorder and simplify the analysis.

\subsubsection{The Sensor, Controller and Actuator}
The sensor is time-driven with a constant sampling period $T$. At each sampling instant, the measurements are encapsulated into a packet and sent to all controllers via the wireless network (or multihop network). The decentralized controllers are event-driven and operate in discrete-time. For a given controller $i$, once the sensor packet arrives, a new control signal is generated and directly sent to the actuator node. When the control signals reach the actuator after the controller-to-actuator delays, the combination of all control signals is used to control the plant.

\subsubsection{The Controlled Plant}
Considering the network-induced delay, the dynamic of the controlled plant is given by the following linear continuous-time model:
\begin{equation}\label{eq:1}
\dot x\left( t \right) = Ax\left( t \right) + \sum\limits_{i = 1}^p {B_i \tilde u_i \left( {t - \tau _i } \right)} ,
\end{equation}
where $x\left( t \right)$ is an $M$-dimensional plant state vector, the actuator input $\tilde u_i \left( t \right)$ from the $i$-th controller is a $K$-dimensional vector, $A$ and $B_i$ are known matrices of appropriate sizes.

Then, the corresponding discrete-time version of (\ref{eq:1}) is given by
\begin{equation}\label{eq:2}
x_{k + 1}  = \Phi x_k  + \sum\limits_{i = 1}^p {\left( {\Gamma _{i,k}^0 \tilde u_{i,k}  + \Gamma _{i,k}^1 \tilde u_{i,k - 1} } \right)} ,
\end{equation}
where $x_k  = x\left( {kT} \right)$, $\Phi  = e^{AT}$, $\Gamma _{i,k}^0  = \int_0^{T - \tau _{i,k} } {e^{As} dsB_i }$, $\Gamma _{i,k}^1  = \int_{T - \tau _{i,k} }^T {e^{As} ds} B_i$, $\tau _{i,k}$ and $\tilde u_{i,k}$ denote the total time delay and the actuator input of the $i$-th controller, respectively, in response to the sensor signal $x_k$.

Considering the packet loss, $\tilde u_{i,k}$ can be expressed by
\begin{equation}\label{eq:3}
\tilde u_{i,k}  = \theta _{i,k} u_{i,k}  + \left( {1 - \theta _{i,k} } \right)u_{i,k - 1} ,
\end{equation}
where $u_{i,k}$ and $\theta _{i,k}$ are the control signal and the value of switch $\theta _i$, respectively, in response to the sensor signal $x_k$.
Eq.~(\ref{eq:3}) indicates that the actuator remains to use the lastest control signal $u_{i,k - 1}$ when $u_{i,k}$ is lost due to the controller-to-actuator loss or not generated due to the sensor-to-controller loss; otherwise, the newly arrived control signal $u_{i,k}$ is used. However, it is possible that multiple consecutive packets are lost. Iteratively using (\ref{eq:3}), the general formulation of $\tilde u_{i,k}$ is given by
\begin{equation}\label{eq:4}
\tilde u_{i,k}  = \theta _{i,k} u_{i,k}  + \sum\limits_{j = 0}^{k - 1} {\left( {\prod\limits_{l = j + 1}^k {\left( {1 - \theta _{i,l} } \right)} } \right)} \theta _{i,j} u_{i,j} ,
\end{equation}
which means that $\tilde u_{i,k}  = u_{i,k - m}$ when $\theta _{i,l}  = 0,\ l = k,\ k - 1, \cdots ,\ k - m + 1$ and $\theta _{i,k - m}  = 1$, i.e., the latest available control signal $u_{i,k - m}$ is used when the $u_{i,k}$ and the previous ($m - 1$) control signals are all lost or not generated.

Substituting (\ref{eq:4}) into (\ref{eq:2}), the discrete-time control process with delays and packet losses can be expressed as
\begin{equation}\label{eq:5}
x_{k + 1}  = \Phi x_k  + \sum\limits_{i = 1}^p {\sum\limits_{j = 0}^k {\beta _{i,k}^j u_{i,k - j} } } ,
\end{equation}
where
\begin{equation}\label{eq:6}
\beta _{i,k}^j  = \left\{ {\begin{array}{*{20}c}
   {\Gamma _{i,k}^0 \theta _{i,k} ,} & {j = 0,}  \\
   {\left[ {\Gamma _{i,k}^0 \left( {1 - \theta _{i,k} } \right) + \Gamma _{i,k}^1 } \right]\theta _{i,k - 1} ,} & {j = 1,}  \\
   {\left[ {\Gamma _{i,k}^0 \prod\limits_{l = k - j + 1}^k {\left( {1 - \theta _{i,l} } \right)}  + \Gamma _{i,k}^1 \prod\limits_{l = k - j + 1}^{k - 1} {\left( {1 - \theta _{i,l} } \right)} } \right]\theta _{i,k - j} ,} & {others,}  \\
\end{array}} \right.
\end{equation}
which depends on the delay $\tau _{i,k}$ and packet losses $\theta _{i,l} ,\ l = k,\ k - 1, \cdots ,\ k - j$.

Note that, there are many works to model the discrete-time control process with the event-driven controller and actuator under packet losses. Most of them directly set the actuator input to be zero when the packet loss happens \cite{IRef15}, which is a typical way to simply the analysis. However, in the real system, the plant state loss and the control signal loss result in no signal (i.e, no event) arriving at the controller and the actuator, respectively. In this case, the actuator input remains the same as the last control signal due to the event-driven controller and actuator. Thus, we formulate the discrete-time control process as (\ref{eq:5}) and (\ref{eq:6}).

\subsection{Problem Formulation}
In this paper, we focus on the case of WSANs with multiple decentralized controllers, and address its optimal state-feedback control problem in the presence of stochastic network-induced delays and packet losses. Using a quadratic cost function \cite{IRef5}\cite{IRef6}\cite{IRef30}-\cite{IRef32}, the design of the optimal state-feedback control is to minimize the cost function, i.e.,
\begin{equation}\label{eq:7}
\begin{split}
 \mathop {\min }\limits_{\left\{ {u_{i,k} } \right\}} \ &{\rm{  }}J_N  = \mathbb{E}\left[ {x_N^T Q_N x_N  + \sum\limits_{k = 0}^{N - 1} {\left( {x_k^T Q_1 x_k  + \sum\limits_{i = 1}^p {u_{i,k}^T R_i u_{i,k} } } \right)} } \right], \\
 s.t.\ &{\rm{   }}x_{k + 1}  = \Phi x_k  + \sum\limits_{i = 1}^p {\sum\limits_{j = 0}^k {\beta _{i,k}^j u_{i,k - j} } } , \\
 \end{split}
\end{equation}
where $\mathbb{E}$ is the expectation operator over the distributions of all delays and packet losses, $N$ is the total number of sampling instants,
$Q_N  \succeq 0$ is a symmetric positive semi-definite weight matrix, $Q_1 \succ 0$ and $R_i \succ 0$ are symmetric positive definite weight matrices.

Since the controllers are event-driven, the current control signals are generated asynchronously and randomly due to their individual stochastic sensor-to-controller delays. Hence, the current control signals of other controllers are not available when one controller makes the control strategy. That is, the global information is not available anymore. In this case, for the decentralized control system, we can reformulate (\ref{eq:7}) as a non-cooperative control problem among $p$ controllers \cite{IRef30}-\cite{IRef32}:
\begin{equation}\label{eq:8}
\begin{split}
 \mathop {\min }\limits_{\left\{ {u_{i,k} } \right\} } \ &{\rm{  }}J_N^i  = \mathbb{E}\left[ {x_N^T Q_N x_N  + \sum\limits_{k = 0}^{N - 1} {\left( {x_k^T Q_1 x_k  + u_{i,k}^T R_i u_{i,k} } \right)} } \right], \\
 s.t.\ &{\rm{   }}x_{k + 1}  = \Phi x_k  + \sum\limits_{i = 1}^p {\sum\limits_{j = 0}^k {\beta _{i,k}^j u_{i,k - j} } } . \\
 \end{split}
\end{equation}
where $J_N^i, \ i = 1,\ 2, \cdots,\ p,$ is the cost function of controller $i$.

\section{State-Feedback Controller Design}
\label{Derivation}
This section derives the optimal state-feedback control for the non-cooperative game in (\ref{eq:8}). We first assume that the current state information and the past control signals are all perfectly known to the controller. The results are extended to the case of the controller with imperfect information caused by the packet loss in the next section.

Due to the individual stochastic sensor-to-controller delays of multiple controllers, the current control signals of controllers are unavailable for each other. However, it is reasonable to use other controllers' past control signals for the design of the control strategy. For a given controller, if the current state information and the past control signals are all assumed to be perfectly known, the general form of the linear control law based on all known information is given by
\begin{equation}\label{eq:9}
u_{i,k}  = A_i^k x_k  + \sum\limits_{m = 1}^p {\sum\limits_{n = 1}^k {\alpha _{i,k}^{m,n} u_{m,k - n} } } ,\ {\rm{  }}i = 1,\ {\rm{ }}2,{\rm{ }} \cdots ,\ {\rm{ }}p,
\end{equation}
where $A_i^k$ and $\alpha _{i,k}^{m,n}$ are coefficient matrices with appropriate sizes. Note that, in (\ref{eq:9}), the perfect information is assumed to be available for controllers, and the effect of the imperfect information caused by the packet loss will be investigated in the next section.

Define
\begin{equation}\label{eq:10}
\begin{split}
 z_k  &= \left[ {\begin{array}{*{20}c}
   {x_k^T } & {\hat u_{k - 1}^T } & {\hat u_{k - 2}^T } &  \cdots  & {\hat u_0^T }  \\
\end{array}} \right]^T  \in \mathbb{R}^{M + pkK}, \\
 \hat u_j  &= \left[ {\begin{array}{*{20}c}
   {u_{1,j}^T } & {u_{2,j}^T } &  \cdots  & {u_{p,j}^T }  \\
\end{array}} \right]^T ,\ {\rm{  }}j = 0,\ {\rm{ }}1, \cdots ,\ {\rm{ }}k - 1. \\
 \end{split}
\end{equation}

Taking controller $i$ as the desired one, substituting $u_{j,k} ,\ j \ne i$ in (\ref{eq:9}) into (\ref{eq:5}), we get
\begin{equation}\label{eq:11}
x_{k + 1}  = \left( {\Phi  + \sum\limits_{l = 1, l \ne i}^p {\beta _{l,k}^0 A_l^k } } \right)x_k  + \beta _{i,k}^0 u_{i,k}  + \sum\limits_{m = 1}^p {\sum\limits_{j = 1}^k {\left( {\beta _{m,k}^j  + \sum\limits_{l = 1, l \ne i}^p {\beta _{l,k}^0 \alpha _{l,k}^{m,j} } } \right)u_{m,k - j} } } .
\end{equation}

Based on (\ref{eq:10}) and (\ref{eq:11}), we can rewrite (\ref{eq:5}) as
\begin{equation}\label{eq:12}
z_{k + 1}  = C_{i,k} z_k  + D_{i,k} u_{i,k} ,
\end{equation}
where the time-varying coefficient matrices are given by
\begin{equation}\label{eq:13}
{C_{i,k}} = \left[ {\begin{array}{*{20}{c}}
   {\Phi  + \sum\limits_{\scriptstyle l = 1 \hfill \atop
  \scriptstyle l \ne i \hfill}^p {\beta _{l,k}^0A_l^k} } & {\gamma _{1,k}^1} & {\gamma _{2,k}^1} &  \cdots  & {\gamma _{p,k}^1} &  \cdots  & {\gamma _{1,k}^k} & {\gamma _{2,k}^k} &  \cdots  & {\gamma _{p,k}^k}  \\
   {A_1^k} & {\alpha _{1,k}^{1,1}} & {\alpha _{1,k}^{2,1}} &  \cdots  & {\alpha _{1,k}^{p,1}} &  \cdots  & {\alpha _{1,k}^{1,k}} & {\alpha _{1,k}^{2,k}} &  \cdots  & {\alpha _{1,k}^{p,k}}  \\
    \vdots  &  \vdots  &  \vdots  &  \cdots  &  \vdots  &  \cdots  &  \vdots  &  \vdots  &  \cdots  &  \vdots   \\
   {A_{i - 1}^k} & {\alpha _{i - 1,k}^{1,1}} & {\alpha _{i - 1,k}^{2,1}} &  \cdots  & {\alpha _{i - 1,k}^{p,1}} &  \cdots  & {\alpha _{i - 1,k}^{1,k}} & {\alpha _{i - 1,k}^{2,k}} &  \cdots  & {\alpha _{i - 1,k}^{p,k}}  \\
   0 & 0 & 0 & 0 & 0 & 0 & 0 & 0 & 0 & 0  \\
   {A_{i + 1}^k} & {\alpha _{i + 1,k}^{1,1}} & {\alpha _{i + 1,k}^{2,1}} &  \cdots  & {\alpha _{i + 1,k}^{p,1}} &  \cdots  & {\alpha _{i + 1,k}^{1,k}} & {\alpha _{i + 1,k}^{2,k}} &  \cdots  & {\alpha _{i + 1,k}^{p,k}}  \\
    \vdots  &  \vdots  &  \vdots  &  \cdots  &  \vdots  &  \cdots  &  \vdots  &  \vdots  &  \cdots  &  \vdots   \\
   {A_p^k} & {\alpha _{p,k}^{1,1}} & {\alpha _{p,k}^{2,1}} &  \cdots  & {\alpha _{p,k}^{p,1}} &  \cdots  & {\alpha _{p,k}^{1,k}} & {\alpha _{p,k}^{2,k}} &  \cdots  & {\alpha _{p,k}^{p,k}}  \\
   0 & {{I_K}} & 0 &  \cdots  & 0 &  \cdots  & 0 & 0 &  \cdots  & 0  \\
   0 & 0 & {{I_K}} &  \cdots  & 0 &  \cdots  & 0 & 0 &  \cdots  & 0  \\
    \vdots  &  \vdots  &  \vdots  &  \ddots  &  \vdots  &  \ddots  &  \vdots  &  \vdots  &  \ddots  &  \vdots   \\
\end{array}} \right],\ {D_{i,k}} = \left[ {\begin{array}{*{20}{c}}
   {\beta _{i,k}^0}  \\
   0  \\
    \vdots   \\
   0  \\
   {I_K^{i + 1}}  \\
   0  \\
    \vdots   \\
\end{array}} \right],
\end{equation}
where $\gamma _{m,k}^j  = \beta _{m,k}^j  + \sum\nolimits_{l = 1,l \ne i}^p {\beta _{l,k}^0 \alpha _{l,k}^{m,j} }$, $0$ is the zero matrix with appropriate size, $I_K$ is the $K \times K$ identity matrix, and $I_K^{i + 1}$ denotes the $\left( {i + 1} \right)$-th block of $D_{i,k}$ as a $K \times K$ identity matrix.

Then, the optimization problem for controller $i$ in (\ref{eq:8}) can be rewritten as
\begin{equation}\label{eq:14}
\begin{split}
 \mathop {\min }\limits_{\left\{ {u_{i,k} } \right\} } \ &{\rm{  }}J_N^i  = \mathbb{E}\left[ {z_N^T \bar Q_N z_N  + \sum\limits_{k = 0}^{N - 1} {\left[ {\left( {\begin{array}{*{20}c}
   {z_k }  \\
   {u_{i,k} }  \\
\end{array}} \right)^T \left( {\begin{array}{*{20}c}
   {\bar Q_1 } & 0  \\
   0 & {R_i }  \\
\end{array}} \right)\left( {\begin{array}{*{20}c}
   {z_k }  \\
   {u_{i,k} }  \\
\end{array}} \right)} \right]} } \right], \\
 s.t.\ &{\rm{   }}z_{k + 1}  = C_{i,k} z_k  + D_{i,k} u_{i,k} , \\
 \end{split}
\end{equation}
where
\begin{equation}\label{eq:15}
\bar Q_N  = \left( {\begin{array}{*{20}c}
   {Q_N } & 0 &  \cdots  & 0  \\
   0 & 0 &  \cdots  & 0  \\
    \vdots  &  \vdots  &  \ddots  &  \vdots   \\
   0 & 0 &  \cdots  & 0  \\
\end{array}} \right),\ {\rm{ }}\bar Q_1  = \left( {\begin{array}{*{20}c}
   {Q_1 } & 0 &  \cdots  & 0  \\
   0 & 0 &  \cdots  & 0  \\
    \vdots  &  \vdots  &  \ddots  &  \vdots   \\
   0 & 0 &  \cdots  & 0  \\
\end{array}} \right).
\end{equation}

Define
\begin{equation}\label{eq:16}
V_L^i  = \mathop {\min }\limits_{\left\{ {u_{i,k} } \right\} } {\rm{ }}\mathbb{E}\left[ {z_N^T \bar Q_N z_N  + \sum\limits_{k = L}^{N - 1} {\left[ {\left( {\begin{array}{*{20}c}
   {z_k }  \\
   {u_{i,k} }  \\
\end{array}} \right)^T \left( {\begin{array}{*{20}c}
   {\bar Q_1 } & 0  \\
   0 & {R_i }  \\
\end{array}} \right)\left( {\begin{array}{*{20}c}
   {z_k }  \\
   {u_{i,k} }  \\
\end{array}} \right)} \right]} } \right].
\end{equation}

Lemma 1 \cite{IRef43}. Assume that the function $l\left( {x,y,u} \right)$ has a unique minimum with respect to $u \in U$ for all $x \in X$ and $y \in Y$. Let $u^0 \left( {x,y} \right)$ denote the value of $u$ for which the minimum is achieved. Then
\[
\mathop {\min }\limits_{u\left( {x,y} \right)} \mathbb{E}\left[ {l\left( {x,y,u} \right)} \right] = \mathbb{E}\left[ {l\left( {x,y,u^0 \left( {x,y} \right)} \right)} \right] = \mathbb{E}\mathop {\min }\limits_u \left\{ {l\left( {x,y,u} \right)} \right\}.
\]

We next derive the expressions for the optimal decentralized controllers.

\subsubsection{L = N} When $L = N$, we have
\begin{equation}\label{eq:17}
V_N^i  = \mathbb{E}\left[ {z_N^T S_{i,N} z_N } \right] ,
\end{equation}
where $S_{i,N}  = \bar Q_N$.

\subsubsection{$L = N-1$}
When $L = N-1$, based on Lemma 1, from (\ref{eq:14}), (\ref{eq:16}) and (\ref{eq:17}), we get
\begin{equation}\label{eq:18}
\begin{split}
 V_{N - 1}^i  &= \mathop {\min }\limits_{u_{i,N - 1} } \mathbb{E}\left\{ {\left. {\left( {\begin{array}{*{20}c}
   {z_{N - 1} }  \\
   {u_{i,N - 1} }  \\
\end{array}} \right)^T \left( {\begin{array}{*{20}c}
   {\bar Q_1 } & 0  \\
   0 & {R_i }  \\
\end{array}} \right)\left( {\begin{array}{*{20}c}
   {z_{N - 1} }  \\
   {u_{i,N - 1} }  \\
\end{array}} \right) + V_N^i } \right|z_{N - 1} } \right\}, \\
 {\rm{      }} &= \mathop {\min }\limits_{u_{i,N - 1} } \mathbb{E}\left\{ {\left. {\left( {\begin{array}{*{20}c}
   {z_{N - 1} }  \\
   {u_{i,N - 1} }  \\
\end{array}} \right)^T \left( {\begin{array}{*{20}c}
   {P_{i,N - 1} } & {\bar P_{i,N - 1}^T }  \\
   {\bar P_{i,N - 1} } & {\tilde P_{i,N - 1} }  \\
\end{array}} \right)\left( {\begin{array}{*{20}c}
   {z_{N - 1} }  \\
   {u_{i,N - 1} }  \\
\end{array}} \right)} \right|z_{N - 1} } \right\}, \\
 {\rm{      }} &= \mathbb{E}\mathop {\min }\limits_{u_{i,N - 1} } \left\{ {\left. {\left( {\begin{array}{*{20}c}
   {z_{N - 1} }  \\
   {u_{i,N - 1} }  \\
\end{array}} \right)^T \left( {\begin{array}{*{20}c}
   {P_{i,N - 1} } & {\bar P_{i,N - 1}^T }  \\
   {\bar P_{i,N - 1} } & {\tilde P_{i,N - 1} }  \\
\end{array}} \right)\left( {\begin{array}{*{20}c}
   {z_{N - 1} }  \\
   {u_{i,N - 1} }  \\
\end{array}} \right)} \right|z_{N - 1} } \right\}, \\
 \end{split}
\end{equation}
where
\begin{equation}\label{eq:19}
\begin{split}
 P_{i,N - 1}  &= \bar Q_1  + \mathbb{E}\left[ {C_{i,N - 1}^T S_{i,N} C_{i,N - 1} } \right], \\
 \bar P_{i,N - 1}  &= \mathbb{E}\left[ {D_{i,N - 1}^T S_{i,N} C_{i,N - 1} } \right], \\
 \tilde P_{i,N - 1}  &= R_i  + \mathbb{E}\left[ {D_{i,N - 1}^T S_{i,N} D_{i,N - 1} } \right]. \\
 \end{split}
\end{equation}

Then, the optimal solution to (\ref{eq:19}) is given by \cite{IRef44}
\begin{equation}\label{eq:20}
u_{i,N - 1}  =  - L_{i,N - 1} z_{N - 1} ,
\end{equation}
where
\begin{equation}\label{eq:21}
\begin{split}
 L_{i,N - 1}  &= \left( {\tilde P_{i,N - 1} } \right)^{ - 1} \bar P_{i,N - 1}  \\
 {\rm{       }} &= \left[ {R_i  + \mathbb{E}\left[ {D_{i,N - 1}^T S_{i,N} D_{i,N - 1} } \right]} \right]^{ - 1} \mathbb{E}\left[ {D_{i,N - 1}^T S_{i,N} C_{i,N - 1} } \right]. \\
 \end{split}
\end{equation}
%

Then, substituting $u_{i,N - 1}$ in (\ref{eq:20}) into (\ref{eq:18}), $V_{N - 1}^i$ is deduced as
\begin{equation}\label{eq:22}
V_{N - 1}^i  = \mathbb{E}\left[ {z_{N - 1}^T S_{i,N - 1} z_{N - 1} } \right],
\end{equation}
where
\begin{equation}\label{eq:23}
S_{i,N - 1}  = \bar Q_1  + \mathbb{E}\left[ {C_{i,N - 1}^T S_{i,N} C_{i,N - 1} } \right] - L_{i,N - 1}^T \left[ {R_i  + \mathbb{E}\left[ {D_{i,N - 1}^T S_{i,N} D_{i,N - 1} } \right]} \right]L_{i,N - 1} .
\end{equation}

\subsubsection{$L = N - 2, \cdots ,\ 1,\ 0$}
When $L = N - 2$, from (\ref{eq:16}) and (\ref{eq:22}), we have
\begin{equation}\label{eq:24}
V_{N - 2}^i  = \mathop {\min }\limits_{u_{i,N - 2} } \mathbb{E}\left[ {\left. {\left( {\begin{array}{*{20}c}
   {z_{N - 2} }  \\
   {u_{i,N - 2} }  \\
\end{array}} \right)^T \left( {\begin{array}{*{20}c}
   {\bar Q_1 } & 0  \\
   0 & {R_i }  \\
\end{array}} \right)\left( {\begin{array}{*{20}c}
   {z_{N - 2} }  \\
   {u_{i,N - 2} }  \\
\end{array}} \right) + V_{N - 1}^i } \right|z_{N - 2} } \right],
\end{equation}
which has the same form as (\ref{eq:18}). Thus, repeat the same process as that for $L = N - 1$, we can derive the optimal state-feedback control law $u_{i,k} ,\ {\rm{  }}k = N - 2, \cdots ,\ {\rm{ 1}},\ {\rm{ }}0$, which is given by
\begin{equation}\label{eq:25}
u_{i,k}  =  - L_{i,k} z_k ,\ {\rm{  }}i = 1,\ {\rm{ }}2{\rm{ }} \cdots ,\ {\rm{ }}p;\ {\rm{ }}k = 0,\ {\rm{ }}1,{\rm{ }} \cdots ,\ {\rm{ }}N - 1,
\end{equation}
where
\begin{equation}\label{eq:26}
\begin{split}
 L_{i,k}  &= \left[ {R_i  + \mathbb{E}\left[ {D_{i,k}^T S_{i,k + 1} D_{i,k} } \right]} \right]^{ - 1} \mathbb{E}\left[ {D_{i,k}^T S_{i,k + 1} C_{i,k} } \right], \\
 S_{i,k}  &= \bar Q_1  + \mathbb{E}\left[ {C_{i,k}^T S_{i,k + 1} C_{i,k} } \right] - L_{i,k}^T \left[ {R_i  + \mathbb{E}\left[ {D_{i,k}^T S_{i,k + 1} D_{i,k} } \right]} \right]L_{i,k} , \\
 \end{split}
\end{equation}
and the corresponding
\begin{equation}\label{eq:27}
V_k^i  = \mathbb{E}\left[ {z_k^T S_{i,k} z_k } \right]. \\
\end{equation}

We observe that the optimal decentralized state-feedback control law $L_{i,k}$ is obtained using a backward recursion based on (\ref{eq:25}) and (\ref{eq:26}).

From (\ref{eq:9}) and (\ref{eq:25}), we get
\begin{equation}\label{eq:28}
L_{i,k}  =  - \left[ {\begin{array}{*{20}c}
   {A_i^k } & {\alpha _{i,k}^{1,1} } & {\alpha _{i,k}^{2,1} } &  \cdots  & {\alpha _{i,k}^{p,1} } &  \cdots  & {\alpha _{i,k}^{1,k} } & {\alpha _{i,k}^{2,k} } &  \cdots  & {\alpha _{i,k}^{p,k} }  \\
\end{array}} \right].
\end{equation}

Combining the expressions of $L_{i,k}$ in (\ref{eq:26}) and (\ref{eq:28}), for a given controller $i,\ {\rm{ }}i = 1,\ {\rm{ 2}}, \cdots ,\ p$, we can derive
\begin{equation}\label{eq:29}
\begin{split}
 &A_i^k  = G_i^{ - 1} \left[ {\left( {\beta _{i,k}^0 } \right)^T S_{i,k + 1}^{1,1} \Phi  + S_{i,k + 1}^{i + 1,1} \Phi  + \sum\limits_{\scriptstyle l = 1 \hfill \atop
  \scriptstyle l \ne i \hfill}^p {Y_{i,k}^l A_l^k } } \right], \\
 &\alpha _{i,k}^{m,n} = G_i^{ - 1}\left[ {{{\left( {\beta _{i,k}^0} \right)}^T}S_{i,k + 1}^{1,1}\beta _{m,k}^n + S_{i,k + 1}^{i + 1,1}\beta _{m,k}^n + \sum\limits_{\scriptstyle l = 1 \hfill \atop
  \scriptstyle l \ne i \hfill}^p {Y_{i,k}^l\alpha _{l,k}^{m,n}}  + {{\left( {\beta _{i,k}^0} \right)}^T}S_{i,k + 1}^{1,pj + m + 1} + S_{i,k + 1}^{i + 1,pj + m + 1}} \right], \\
 &m = 1,\ {\rm{ }}2, \cdots ,\ {\rm{ }}p;\ {\rm{ }}n = 1,\ {\rm{ }}2, \cdots ,\ {\rm{ }}k - 1, \\
 \end{split}
\end{equation}
where
\begin{equation}
\begin{split}
 G_i  &= R_i  + \mathbb{E}\left[ {D_{i,k}^T S_{i,k + 1} D_{i,k} } \right], \\
 Y_{i,k}^l  &= \left( {\beta _{i,k}^0 } \right)^T S_{i,k + 1}^{1,1} \beta _{l,k}^0  + S_{i,k + 1}^{i + 1,1} \beta _{l,k}^0  + \left( {\beta _{i,k}^0 } \right)^T S_{i,k + 1}^{1,l + 1}  + S_{i,k + 1}^{i + 1,l + 1} , \\
 \end{split}
\end{equation}
and $S_{i,k}^{m,n}$ denotes the ($m,n$)-th block of matrix $S_{i,k}$.

It can be seen that all matrix equations in (\ref{eq:29}) are linear. We can easily calculate the values of the coefficient matrices $A_i^k$ and $\alpha _{i,k}^{m,n}$, $i = 1,\ {\rm{ }}2, \cdots ,\ {\rm{ }}p$, $m = 1,\ {\rm{ }}2, \cdots ,\ {\rm{ }}p$, $n = 1,\ {\rm{ }}2, \cdots ,\ {\rm{ }}k$. Then we can obtain the optimal control strategy from (\ref{eq:25}) and (\ref{eq:28}). For example, for the case of two decentralized controllers, i.e., $p=2$, the solutions for $A_i^k$ and $\alpha _{i,k}^{m,n}$, $i = 1,\ {\rm{ }}2$, $m = 1,\ {\rm{ }}2$, $n = 1,\ {\rm{ }}2, \cdots ,\ {\rm{ }}k$, are given as follows:
\begin{equation}
\begin{split}
 &A_1^k  = \left[ {I - a_2^1 \left( k \right)a_2^2 \left( k \right)} \right]^{ - 1} \left[ {a_2^1 \left( k \right)a_1^2 \left( k \right) - a_1^1 \left( k \right)} \right], \\
 &A_2^k  = \left[ {I - a_2^2 \left( k \right)a_2^1 \left( k \right)} \right]^{ - 1} \left[ {a_2^2 \left( k \right)a_1^1 \left( k \right) - a_1^2 \left( k \right)} \right], \\
 &\alpha _{1,k}^{m,n}  = \left[ {I - b_{2,n}^{1,m} \left( k \right)b_{2,n}^{2,m} \left( k \right)} \right]^{ - 1} \left[ {b_{2,n}^{1,m} \left( k \right)b_{1,n}^{2,m} \left( k \right) - b_{1,n}^{1,m} \left( k \right)} \right], \\
 &\alpha _{2,k}^{m,n}  = \left[ {I - b_{2,n}^{2,m} \left( k \right)b_{2,n}^{1,m} \left( k \right)} \right]^{ - 1} \left[ {b_{2,n}^{2,m} \left( k \right)b_{1,n}^{1,m} \left( k \right) - b_{1,n}^{2,m} \left( k \right)} \right], \\
 \end{split}
\end{equation}
where, for $i=1,\ 2$,
\begin{equation}
\begin{split}
 &a_2^i\left( k \right) = b_{2,n}^{i,m}\left( k \right) = G_i^{ - 1}\mathbb{E}\left[ {{{\left( {\beta _{i,k}^0} \right)}^T}S_{i,k + 1}^{1,1}\beta _{3 - i,k}^0 + S_{i,k + 1}^{i + 1,1}\beta _{3 - i,k}^0 + {{\left( {\beta _{i,k}^0} \right)}^T}S_{i,k + 1}^{1,4 - i} + S_{i,k + 1}^{i + 1,4 - i}} \right], \\
 &a_1^i\left( k \right) = G_i^{ - 1}\mathbb{E}\left[ {{{\left( {\beta _{i,k}^0} \right)}^T}S_{i,k + 1}^{1,1}\Phi  + S_{i,k + 1}^{i + 1,1}\Phi } \right], \\
 &b_{1,n}^{i,m}\left( k \right) = G_i^{ - 1}\mathbb{E}\left[ {{{\left( {\beta _{i,k}^0} \right)}^T}S_{i,k + 1}^{1,1}\beta _{m,k}^n + S_{i,k + 1}^{i + 1,1}\beta _{m,k}^n + {{\left( {\beta _{i,k}^0} \right)}^T}S_{i,k + 1}^{1,2j + m + 1} + S_{i,k + 1}^{i + 1,2j + m + 1}} \right]. \\
 \end{split}
\end{equation}

\begin{table}
\caption{Optimal state-feedback control algorithm with decentralized controllers.}
\label{tab1}       
\begin{tabular}{ll}
\hline\noalign{\smallskip}
    \;\textbf {Offline}: \\
    \;1: Initialize $S_{i,N}  = \bar Q_N ,\ i = 1,\ 2, \cdots ,\ p$.\\
    \;2:  \textbf {for $k = N - 1: - 1:0$ do} \\
    \;3:\ \ \ Calculate $A_i^k$ and $\alpha _{i,k}^{m,n}$, $i = 1,\ {\rm{ }}2, \cdots ,\ {\rm{ }}p$; $m = 1,\ {\rm{ }}2, \cdots ,\ {\rm{ }}p$; $n = 1,\ {\rm{ }}2, \cdots ,\ {\rm{ }}k$ using (\ref{eq:29}). \\
    \;\ \ \ \ \ Calculate $L_{i,k} ,\ i = 1,\ 2, \cdots ,\ p$ from (\ref{eq:28}).\\
    \;4: \textbf{end for} \\
    \;\textbf {On-line}: \\
    \;For a given controller $i,\ i = 1,\ 2, \cdots ,\ p$. \\
    \;1: Initialize $x_0$ and $u_{i,k} = 0,\ {\rm{ }}k < 0$. \\
    \;2:  \textbf {for $k = 0: 1:N - 1$ do} \\
    \;3:\ \ \ \textbf {If (packet lost) } \\
    \:\ \ \ \ \ \ \ Do nothing.\\
    \:\ \ \ \ \ \textbf {else} \\
    \:\ \ \ \ \ \ \ Use $L_{i,k}$ and $z_k$ to compute $u_{i,k}$ in (\ref{eq:25}).\\
    \;4:\ \ \ \textbf {end if} \\
    \;\ \ \ \ \ Exchange control signals $u_{i,k}$ among controllers \\
    \;5: \textbf{end for} \\
\noalign{\smallskip}\hline
\end{tabular}\\
\end{table}

From (\ref{eq:25}) and (\ref{eq:28}), we observe that the optimal state-feedback control
law is linear with the current plant state and all past control signals of decentralized controllers, which
allows each controller to generate the control signal using its local information. Finally,
the procedure for computing the optimal decentralized state-feedback controllers is summarized in Table~\ref{tab1}.

Compared with the derivation of the optimal state-feedback control law for the NCS with single controller, the case of multiple decentralized controllers considered in the paper are much more involved: First, the general form of the optimal linear state-feedback control law should be considered in (\ref{eq:9}) that is a function of current plant state and all previous control signals. Second, the interaction among decentralized controllers should be investigated (see (\ref{eq:11}) and (\ref{eq:12})) and the corresponding partial cost function $V_k^i$ has to be proved to be a quadratic function of $z_k$ (see (\ref{eq:17}), (\ref{eq:22}) and (\ref{eq:27})). Third, the mathematical induction must be employed for the derivation of the optimal control law.

Note that, the optimal state feedback control law in (\ref{eq:25}) is derived for the finite-horizon case. However, the result can be easily extended to the infinite-horizon case $N \to \infty $. From (\ref{eq:26}), if $N$ is large enough, $S_{i,k}$ and $L_{i,k}$ will converge to be constant values ${\bar S}_i$ and ${\bar L}_i$, respectively, using a backward recursion. Thus, ${\bar L}_i$ can be considered as the optimal state feedback control strategy for the infinite-horizon case, since $N \to \infty $ makes all control strategy $L_{i,k} \to {\bar L}_i$ when $k$ is finite. In the design in the Table 1, we can choose a $N$ large enough, and derive the convergent value of ${\bar L}_i$ using the\ ``Off-line"
algorithm. Then, ${\bar L}_i$ is used as the optimal control strategy of the $i$-th controller in the\ ``On-line" part. On the other hand, from (\ref{eq:25}) and (\ref{eq:26}), a drawback with the optimal state-feedback control law is the complicated matrix $L_{i,k}$, since $L_{i,k}$ has to be calculated for each sampling period. An alternative to reduce the complexity of the optimal control law is to use the ${\bar L}_i$ as the suboptimal solution for the finite-horizon case.

\section{The Effect of Imperfect Information Caused by Packet Losses}
\label{extension}
In this section, we extend the results in Section~\ref{Derivation} to the case when the state information and the past control signals of all controllers are not perfectly known. Some detailed derivations will be omitted since they are similar to those in Section~\ref{Derivation}.

From Fig.~\ref{Fig1} and Fig.~\ref{Fig2}, we observe that, in the transmission among controllers through the multihop network, the packet containing the control signal might be dropped or lost. Also, the packet loss exists in the transmission from the sensor to the controller. Considering the packet loss, the linear control law in (\ref{eq:9}) can be rewritten as
\begin{equation}\label{eq:33}
{u_{i,k}} = A_i^k{\bar x_k} + \sum\limits_{m = 1}^p {\sum\limits_{n = 1}^k {\alpha _{i,k}^{m,n}{{\bar u}_{m,k - n}}} } ,\ {\rm{  }}i = 1,\ {\rm{ }}2,{\rm{ }} \cdots ,\ {\rm{ }}p,
\end{equation}
where ${{\bar x}_k}{\rm{ = }}\theta _{i,k}^{sc}{x_k}$ and the value of switch $\theta _{i,k}^{sc}$ denotes the sensor-to-controller loss, ${{\bar u}_{m,k - n}} = \theta _{i,k}^{m,n}{u_{m,k - n}}$ and the value of switch $\theta _{i,k}^{m,n}$ denotes the packet loss from the controller $m$ to the controller $i$. Note that, if the packet is lost, the value of switch is zero, which means there is no signal received by controller $i$.

Similar to (\ref{eq:12}), we have
\begin{equation}
{z_{k + 1}} = {\bar C_{i,k}}{z_k} + {D_{i,k}}{u_{i,k}},
\end{equation}
where
\begin{equation}
{{\bar C}_{i,k}} = \left[ {\begin{array}{*{20}{c}}
   {\Phi  + \sum\limits_{\scriptstyle l = 1 \hfill \atop
  \scriptstyle l \ne i \hfill}^p {\theta _{l,k}^{sc}\beta _{l,k}^0A_l^k} } & {\bar \gamma _{1,k}^1} & {\bar \gamma _{2,k}^1} &  \cdots  & {\bar \gamma _{p,k}^1} &  \cdots  & {\bar \gamma _{1,k}^k} & {\bar \gamma _{2,k}^k} &  \cdots  & {\bar \gamma _{p,k}^k}  \\
   {\theta _{1,k}^{sc}A_1^k} & {\bar \alpha _{1,k}^{1,1}} & {\bar \alpha _{1,k}^{2,1}} &  \cdots  & {\bar \alpha _{1,k}^{p,1}} &  \cdots  & {\bar \alpha _{1,k}^{1,k}} & {\bar \alpha _{1,k}^{2,k}} &  \cdots  & {\bar \alpha _{1,k}^{p,k}}  \\
    \vdots  &  \vdots  &  \vdots  &  \cdots  &  \vdots  &  \cdots  &  \vdots  &  \vdots  &  \cdots  &  \vdots   \\
   {\theta _{i - 1,k}^{sc}A_{i - 1}^k} & {\bar \alpha _{i - 1,k}^{1,1}} & {\bar \alpha _{i - 1,k}^{2,1}} &  \cdots  & {\bar \alpha _{i - 1,k}^{p,1}} &  \cdots  & {\bar \alpha _{i - 1,k}^{1,k}} & {\bar \alpha _{i - 1,k}^{2,k}} &  \cdots  & {\bar \alpha _{i - 1,k}^{p,k}}  \\
   0 & 0 & 0 & 0 & 0 & 0 & 0 & 0 & 0 & 0  \\
   {\theta _{i + 1,k}^{sc}A_{i + 1}^k} & {\bar \alpha _{i + 1,k}^{1,1}} & {\bar \alpha _{i + 1,k}^{2,1}} &  \cdots  & {\bar \alpha _{i + 1,k}^{p,1}} &  \cdots  & {\bar \alpha _{i + 1,k}^{1,k}} & {\bar \alpha _{i + 1,k}^{2,k}} &  \cdots  & {\bar \alpha _{i + 1,k}^{p,k}}  \\
    \vdots  &  \vdots  &  \vdots  &  \cdots  &  \vdots  &  \cdots  &  \vdots  &  \vdots  &  \cdots  &  \vdots   \\
   {\theta _{p,k}^{sc}A_p^k} & {\bar \alpha _{p,k}^{1,1}} & {\bar \alpha _{p,k}^{2,1}} &  \cdots  & {\bar \alpha _{p,k}^{p,1}} &  \cdots  & {\bar \alpha _{p,k}^{1,k}} & {\bar \alpha _{p,k}^{2,k}} &  \cdots  & {\bar \alpha _{p,k}^{p,k}}  \\
   0 & {{I_K}} & 0 &  \cdots  & 0 &  \cdots  & 0 & 0 &  \cdots  & 0  \\
   0 & 0 & {{I_K}} &  \cdots  & 0 &  \cdots  & 0 & 0 &  \cdots  & 0  \\
    \vdots  &  \vdots  &  \vdots  &  \ddots  &  \vdots  &  \ddots  &  \vdots  &  \vdots  &  \ddots  &  \vdots   \\
\end{array}} \right],
\end{equation}
and $\bar \gamma _{m,k}^j = \beta _{m,k}^j + \sum\nolimits_{l = 1,l \ne i}^p {\theta _{l,k}^{m,j}\beta _{l,k}^0\alpha _{l,k}^{m,j}}$, $\bar \alpha _{i,k}^{m,j} = \theta _{i,k}^{m,j}\alpha _{i,k}^{m,j}$, $m = 1,\ 2, \cdots ,\ p,\ j = 1,\ 2, \cdots ,\ k$.

Repeat the derivation from (\ref{eq:14}) to (\ref{eq:25}), the optimal state-feedback control law is obtained as
\begin{equation}
{u_{i,k}} =  - {\bar L_{i,k}}{z_k},\ {\rm{  }}i = 1,\ {\rm{ }}2{\rm{ }} \cdots ,\ {\rm{ }}p;\ {\rm{ }}k = 0,\ {\rm{ }}1,{\rm{ }} \cdots ,\ {\rm{ }}N - 1,
\end{equation}
where
\begin{equation}
\begin{split}
 &{\bar L_{i,k}} = {\left[ {{R_i} + \mathbb{E}\left[ {D_{i,k}^T{\bar S_{i,k + 1}}{D_{i,k}}} \right]} \right]^{ - 1}}\mathbb{E}\left[ {D_{i,k}^T{\bar S_{i,k + 1}}{{\bar C}_{i,k}}} \right], \\
 &{\bar S_{i,k}} = {{\bar Q}_1} + \mathbb{E}\left[ {\bar C_{i,k}^T{\bar S_{i,k + 1}}{{\bar C}_{i,k}}} \right] - \bar L_{i,k}^T\left[ {{R_i} + \mathbb{E}\left[ {D_{i,k}^T{\bar S_{i,k + 1}}{D_{i,k}}} \right]} \right]{\bar L_{i,k}}. \\
 \end{split}
\end{equation}

Then, similar to (\ref{eq:29}), for a given controller $i$, we derive
\begin{equation}\label{eq:38}
\begin{split}
 &A_i^k = {\left[ {\mathbb{E}\left( {\theta _{l,k}^{sc}} \right){{\bar G}_i}} \right]^{ - 1}}\left[ {{{\left( {\beta _{i,k}^0} \right)}^T}\bar S_{i,k + 1}^{1,1}\Phi  + \bar S_{i,k + 1}^{i + 1,1}\Phi  + \sum\limits_{\scriptstyle l = 1 \hfill \atop
  \scriptstyle l \ne i \hfill}^p {\mathbb{E}\left( {\theta _{l,k}^{sc}} \right)\bar Y_{i,k}^lA_l^k} } \right], \\
 &\alpha _{i,k}^{m,n} = {\left[ {\mathbb{E}\left( {\theta _{i,k}^{m,n}} \right){{\bar G}_i}} \right]^{ - 1}}\left[ {{{\left( {\beta _{i,k}^0} \right)}^T}\bar S_{i,k + 1}^{1,1}\beta _{m,k}^n + \bar S_{i,k + 1}^{i + 1,1}\beta _{m,k}^n + } \right.\sum\limits_{\scriptstyle l = 1 \hfill \atop
  \scriptstyle l \ne i \hfill}^p {\mathbb{E}\left( {\theta _{l,k}^{m,n}} \right)\bar Y_{i,k}^l\alpha _{l,k}^{m,n}}  \\
 &\quad \quad \quad \quad \left. {{\rm{                }} + {{\left( {\beta _{i,k}^0} \right)}^T}\bar S_{i,k + 1}^{1,pj + m + 1} + \bar S_{i,k + 1}^{i + 1,pj + m + 1}} \right], \\
 &m = 1,\ {\rm{ }}2, \cdots ,\ {\rm{ }}p;\ {\rm{ }}n = 1,\ {\rm{ }}2, \cdots ,\ {\rm{ }}k - 1, \\
 \end{split}
\end{equation}
where
\begin{equation}
\begin{split}
 &{{\bar G}_i} = {R_i} + \mathbb{E}\left[ {D_{i,k}^T{{\bar S}_{i,k + 1}}{D_{i,k}}} \right], \\
 &\bar Y_{i,k}^l = {\left( {\beta _{i,k}^0} \right)^T}\bar S_{i,k + 1}^{1,1}\beta _{l,k}^0 + \bar S_{i,k + 1}^{i + 1,1}\beta _{l,k}^0 + {\left( {\beta _{i,k}^0} \right)^T}\bar S_{i,k + 1}^{1,l + 1} + \bar S_{i,k + 1}^{i + 1,l + 1}. \\
 \end{split}
\end{equation}

We observe that all equations in (\ref{eq:38}) are linear. Similarly, we can easily calculate the values of $A_i^k$ and $\alpha _{i,k}^{m,n}$ to obtain the optimal control strategy. Note that, if all state information and past control signals of controllers are perfectly known, i.e., $\mathbb{E}\left( {\theta _{l,k}^{sc}} \right) = 1$ and $\mathbb{E}\left( {\theta _{i,k}^{m,n}} \right) = 1$, the result in (\ref{eq:38}) can be reduced to be that in (\ref{eq:29}).

\section{Simulation Results}
\label{Simulation}
In this section, we provide simulation studies on the performance of the proposed stochastic optimal decentralized control algorithms in the NCS with stochastic delays and packet losses. First we consider a genetic control system, and then a power-grid application is investigated. In the simulations, we focus on the case of two decentralized controllers.

\subsection{A Generic System}
First, we consider an NCS as in \cite{IRef5}, and the parameters are set as follows.
\begin{equation}
\begin{split}
 &A = \left[ {\begin{array}{*{20}c}
   0 & 1  \\
   { - 3} & { - 4}  \\
\end{array}} \right],\ {\rm{ }}B_1  = B_2  = \left[ {\begin{array}{*{20}c}
   0  \\
   1  \\
\end{array}} \right], \\
 &Q_N  = Q_1  = 80 \times \left[ {\begin{array}{*{20}c}
   {35} & {\sqrt {35} }  \\
   {\sqrt {35} } & 1  \\
\end{array}} \right], \\
&{\rm{ }}R_1  = R_2  = 10, \\
 \end{split}
\end{equation}
and the sampling period and the length of sampling period are chosen as $T=0.05$ and $N = 50$, respectively.

\begin{figure}[h]
  \begin{center}
  \includegraphics[width=0.65\textwidth]{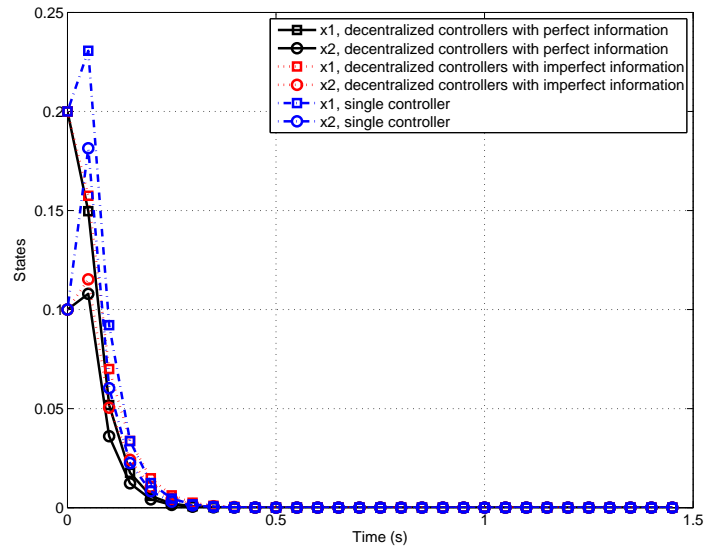}
  \caption{The comparison of plant state responses when $\alpha = 1$ (first 30 sampling periods).} \label{Fig3}
  \end{center}
\end{figure}

\begin{figure}[h]
  \begin{center}
  \includegraphics[width=0.65\textwidth]{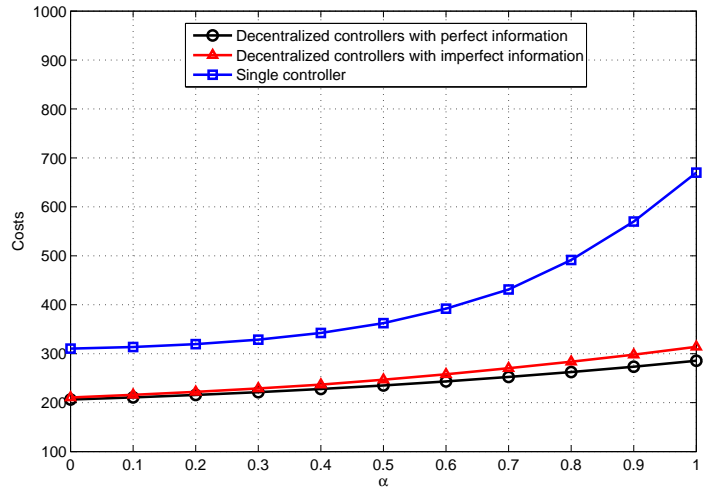}
  \caption{Performance comparison in the generic system.} \label{Fig4}
  \end{center}
\end{figure}

It is assumed that the sensor-to-controller, controller-to-actuator, and controller-to-controller packet losses follow the same Bernoulli distribution with $p = 0.9$, the initial value of the plant state is $x_0  = \left[ {\begin{array}{*{20}c}
   { 0.2} & { 0.1}  \\
\end{array}} \right]^T$, and the network-induced delay is uniform in $\left[ {0,\alpha T} \right],\ 0 \le \alpha  \le 1$. Fig.~\ref{Fig3} and Fig.~\ref{Fig4} show the performance comparison for three cases: decentralized controllers with perfect information, decentralized controllers with imperfect information caused by packet losses, and the single-controller case. 

In Fig.~\ref{Fig3}, $x_j,\ j=1,\ 2$ denotes the $i$-th dimension of the plant state, and it can been seen that the decentralized controllers can make the plant state converge faster than the case of single controller while ensuring the NCS stability in the presence of stochastic delays and packet losses. Fig.~\ref{Fig4} shows the cost function can be significantly reduced by multiple controllers, which indicates that the decentralized controllers in the NCS is an effective way to improve the system performance and stability. In addition, we observe that the imperfect information caused by the packet loss introduces certain performance degradation. We also directly apply the optimal control law of single-controller case to each controller for the multiple-controller case, and obtain that it results the NCS with decentralized multiple be unstable, which means that the single-controller algorithm can not be directly used to the multiple-controller case.

\subsection{Load Frequency Control in Power Grid}
\begin{figure}[h]
  \begin{center}
  \includegraphics[width=0.65\textwidth]{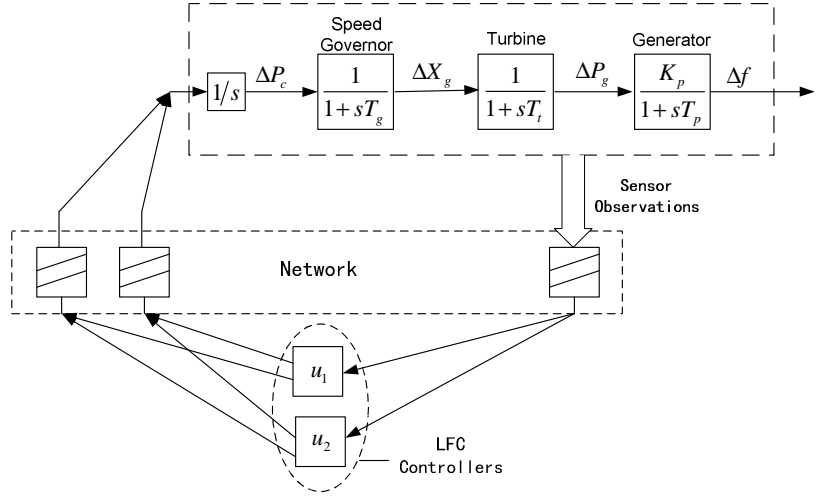}
  \caption{Block diagram of an LFC system for power grid.} \label{Fig5}
  \end{center}
\end{figure}

In this subsection, the application of the proposed stochastic optimal decentralized control scheme to LFC system in power grid \cite{IRef45}\cite{IRef46} is investigated. The typical LFC system is composed of speed governor, turbine, generator and LFC controllers, which is illustrated in Fig.~\ref{Fig5}. The objective is to optimally adjust speed $u_i$ to keep the frequency deviation $\Delta f$ within prescribed limits. The deviations of the generator-turbine-governor system can be represented by two time constants, $T_t$ of the turbine and $T_g$ of the governor. The generator response is considered to be instantaneous in comparison with the time constants of turbine and governor, which can be written as
\begin{equation}\label{eq:47}
\begin{split}
 \frac{d}{{dt}}\Delta P_g  &=  - \frac{1}{{T_t }}\Delta P_g  + \frac{1}{{T_t }}\Delta X_g , \\
 \frac{d}{{dt}}\Delta X_g  &=  - \frac{1}{{T_g }}\Delta X_g  + \frac{1}{{T_g }}\Delta P_c , \\
 \end{split}
\end{equation}
where $\Delta P_g$, $\Delta X_g$ and $\Delta P_c$ are the deviations of generator mechanical output, valve position and generator output, respectively.

The deviation of frequency $\Delta f$ is given by
\begin{equation}\label{eq:48}
\frac{d}{{dt}}\Delta f =  - \frac{1}{{T_p }}\Delta f + \frac{{K_p }}{{T_p }}\Delta P_g ,
\end{equation}
where $K_p$ is the electric system gain, and $T_p$ is the electric system time constant.

The system state vector can be defined as
\begin{equation}
x\left( t \right) = \left[ {\begin{array}{*{20}c}
   {\Delta P_c } & {\Delta f} & {\Delta P_g } & {\Delta X_g }  \\
\end{array}} \right]^T .
\end{equation}
Then, the linear dynamic control model can be described as
\begin{equation}
\dot x\left( t \right) = Ax\left( t \right) + B_1 u_1 \left( {t - \tau _1 } \right) + B_2 u_2 \left( {t - \tau _2 } \right),
\end{equation}
where $B_1 = B_2 = \left[ {\begin{array}{*{20}c}
   1 & 1 & 0 & 0  \\
\end{array}} \right]$, and based on (\ref{eq:47}) and (\ref{eq:48}), $A$ is given by
\begin{equation}\label{eq:51}
A = \left[ {\begin{array}{*{20}c}
   0 & 0 & 0 & 0  \\
   0 & {{{ - 1} \mathord{\left/
 {\vphantom {{ - 1} {T_p }}} \right.
 \kern-\nulldelimiterspace} {T_p }}} & {{{K_p } \mathord{\left/
 {\vphantom {{K_p } {T_p }}} \right.
 \kern-\nulldelimiterspace} {T_p }}} & 0  \\
   0 & 0 & {{{ - 1} \mathord{\left/
 {\vphantom {{ - 1} {T_t }}} \right.
 \kern-\nulldelimiterspace} {T_t }}} & {{1 \mathord{\left/
 {\vphantom {1 {T_t }}} \right.
 \kern-\nulldelimiterspace} {T_t }}}  \\
   {{1 \mathord{\left/
 {\vphantom {1 {T_g }}} \right.
 \kern-\nulldelimiterspace} {T_g }}} & 0 & 0 & {{{ - 1} \mathord{\left/
 {\vphantom {{ - 1} {T_g }}} \right.
 \kern-\nulldelimiterspace} {T_g }}}  \\
\end{array}} \right],
\end{equation}

In the simulation, we set the initial value of the plant state $x_0  = \left[ {\begin{array}{*{20}c}
   { 0.25} & { 0.15} & { 0.2} & { 0.1}  \\
\end{array}} \right]^T$, $K_p  = 1,\ {\rm{ }}T_p  = 0.2,\ {\rm{ }}T_t  = 0.3,\ {\rm{ }}T_g  = 0.08$, and
\begin{equation}\label{eq:52}
 Q_N  = Q_1  = \left[ {\begin{array}{*{20}c}
   1 & 0 & 0 & 0  \\
   0 & 1 & 0 & 0  \\
   0 & 0 & 1 & 0  \\
   0 & 0 & 0 & 1  \\
\end{array}} \right],\ R_1  = R_2  = 1, \\
\end{equation}
and the parameters for delays and packet losses are chosen as the same as in the genetic system.

\begin{figure}[h]
  \begin{center}
  \includegraphics[width=0.65\textwidth]{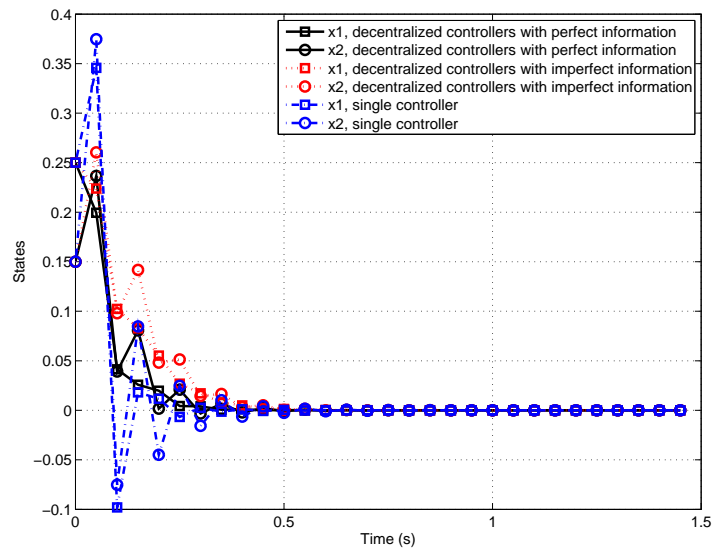}
  \caption{The comparison of plant state responses when $\alpha = 1$ (first two dimensions).} \label{Fig6}
  \end{center}
\end{figure}

\begin{figure}[h]
  \begin{center}
  \includegraphics[width=0.65\textwidth]{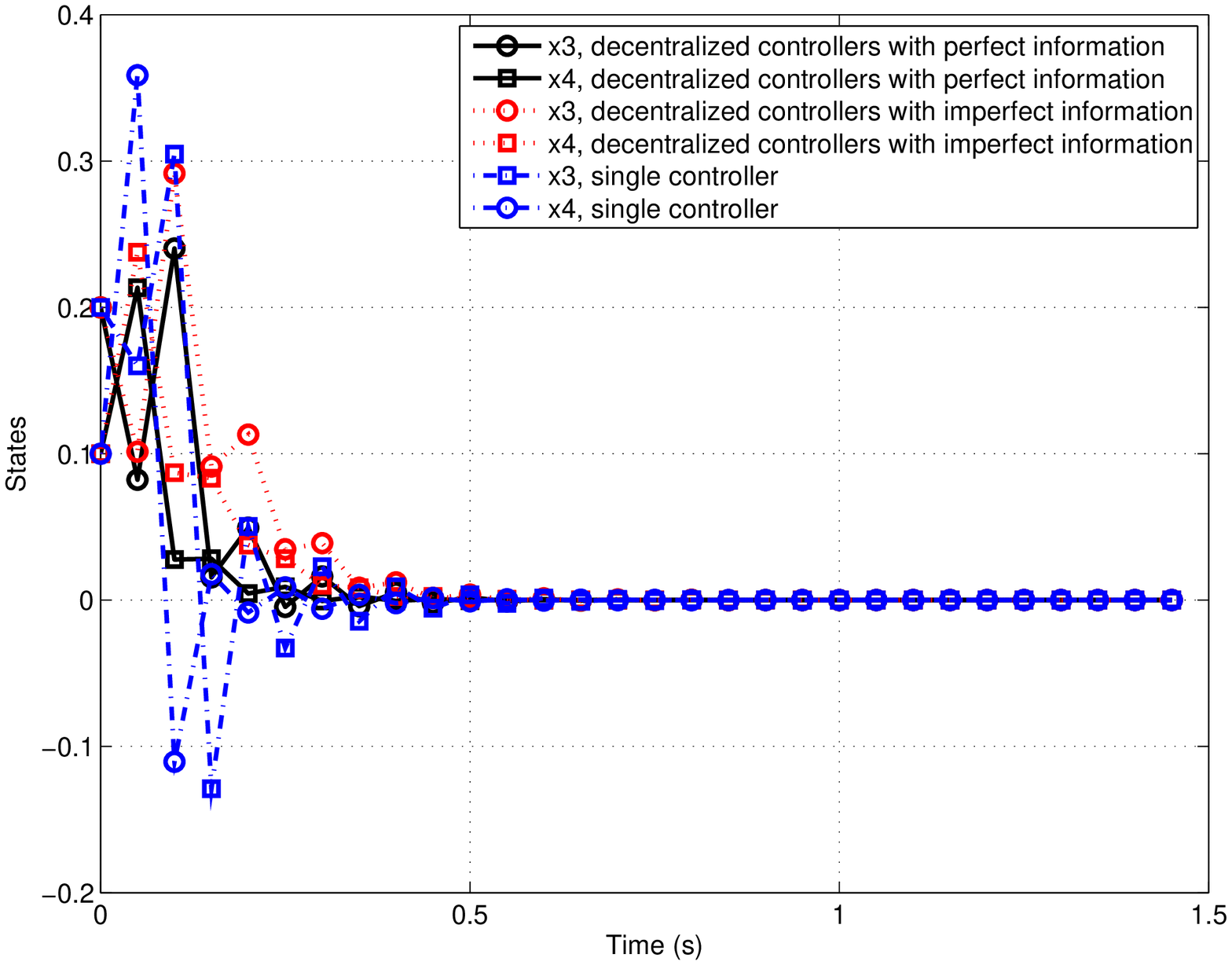}
  \caption{The comparison of plant state responses when $\alpha = 1$ (last two dimensions).} \label{Fig7}
  \end{center}
\end{figure}

\begin{figure}[h]
  \begin{center}
  \includegraphics[width=0.65\textwidth]{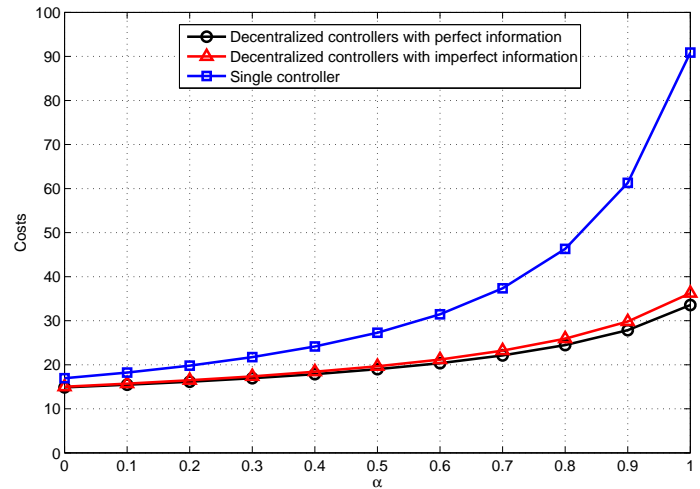}
  \caption{Performance comparison in the LFC system.} \label{Fig8}
  \end{center}
\end{figure}

 From Fig.~\ref{Fig6} - Fig.~\ref{Fig8}, the system performance comparison are also shown for the LFC application. The results are similar to those in the generic control system, and the proposed optimal decentralized control law significantly outperforms the single controller scheme.

\section{Conclusions}
\label{Conclusions}
We have considered the design of the stochastic optimal state-feedback control for NCSs with decentralized controllers in the presence of stochastic delays and packet losses. In particular, the optimization problem is formulated as a non-cooperative LQ game, for which the optimal control solution is derived for decentralized controllers. We have investigated the performance of the proposed algorithm in a genetic control system, as well as a load frequency control system for power grid.

\ifCLASSOPTIONcaptionsoff
  \newpage
\fi

\newpage


\begin{thebibliography}{1}

\bibitem{IRef1}
ISO New England Inc., \emph{Overview of the Smart Grid: Policies, Initiatives and Needs}, \relax Feb. 17, 2009.

\bibitem{IRef2}
S. Chai, G. P. Liu, D. Rees, and Y. Xia,\ ``Design and practical implementation of internet-based predictive control of a servo system," \emph{IEEE Trans. Control Syst. Technol.},
\relax vol. 16, no. 1, pp. 158-168, Jan. 2008.

\bibitem{IRef3}
F. Bullo, J. Cortes and S. Martinez, \emph{Distributed Control of Robotic Networks: A Mathematical Approach to Motion Coordination Algorithms}, \relax Princeton University Press, Princeton, NJ, 2009.

%
%
%
%

\bibitem{IRef4}
W. Zhang, M. S. Branicky, and S. M. Phillips,\ ``Stability of networked control systems,"
\emph{IEEE Control Syst. Mag.}, \relax vol. 21, no. 1, pp. 84-99, Feb. 2001.

\bibitem{IRef5}
J. Nilsson, B. Bernhardsson, and B. Wittenmark,\ ``Stochastic analysis and control of real-time systems with random time delays,"
\emph{Automatica}, \relax vol. 34, no 1, pp. 57-64, Jan. 1998.

\bibitem{IRef6}
S. Hu and W. Zhu,\ ``Stochastic optimal control and analysis of stability of networked control systems with long delay,"
\emph{Automatica}, \relax vol. 39, pp. 1877-1884, Nov. 2003.

\bibitem{IRef7}
J. Hespanha, P. Naghshtabrizi, and Y. Xu,\ ``A survey of recent results in networked control systems,"
\emph{Proc. IEEE}, \relax vol. 95, no. 1, pp. 138-162, Jan. 2007.

\bibitem{IRef8}
L. Zhang, H. Gao, and O. Kaynak,\ ``Network-induced constraints in networked control systems - A survey,"
\emph{IEEE Trans. Ind. Electron.}, \relax vol. 9, no. 1, pp. 403-416, Feb. 2013.

\bibitem{IRef9}
B. Sinopoli, L. Schenato, M. Franceschetti, K. Poolla, M. I. Jordan, and S. S. Sastry,\ ``Kalman filtering with intermittent observations,"
\emph{IEEE Trans. Automat. Contr.}, \relax vol. 49, no. 9, pp. 1453-1464, Sep. 2004.




\bibitem{IRef13}
D. Quevedo and D. Nesic,\ ``Input-to-state stability of packetized predictive control over unreliable networks affected by packet-dropouts,"
\emph{IEEE Trans. Autom. Control}, \relax vol. 56, no. 2, pp. 370-375, Feb. 2011.

\bibitem{IRef14}
C. N. Hadjicostis and R. Touri,\ ``Feedback control utilizing packet dropping network links,"
\emph{in Proc. 41th Conf. Decision and Contr.}, \relax vol. 2, pp. 1205-1210, 2002.
%
%
%

\bibitem{IRef15}
H. Xu, S. Jagannathan, and F. L. Lewis,\ ``Stochastic optimal control of unknown linear networked control system in the presence of random delays and packet losses,"
\emph{Automatica}, \relax vol. 48, pp. 1017-1030, June 2012.

\bibitem{IRef16}
M.~Tabbara, D.~Nesic, and A.~Teel,\ ``Stability of wireless and wireline networked control systems,"
\emph{IEEE Trans. Automat. Contr.}, \relax vol. 52, no. 9, pp. 1615-1630, Sep. 2007.


\bibitem{IRef18}
Hart Communication Foundation. [Online]. {\url{http://www.hartcomm.org/}}

\bibitem{IRef19}
ISA 100 WIRELESS. [Online]. {\url{http://www.isa100wci.org/}}

\bibitem{IRef20}
C. L. Robinson and P. R. Kumar,\ ``Optimizing controller location in networked control systems with packet drops,"
\emph{IEEE J. Sel. Areas Commun.}, \relax vol. 26, no. 4, pp. 661-671, May 2008.

\bibitem{IRef21}
G. C. Goodwin, D. E. Quevedo, and E. I. Silva,\ ``Architectures and coder design for networked control systems,"
\emph{Automatica}, \relax vol. 44, no. 1, pp. 248-257, Jan. 2008.

\bibitem{IRef22}
D. E. Quevedo, K. H. Johansson, A. Ahl¨¦n, and I. Jurado,\ ``Adaptive controller placement for wireless sensor-actuator networks with erasure channels,"
\emph{Automatica}, \relax vol. 49, no. 11, pp. 3458-3466, Nov. 2013.

\bibitem{IRef23}
V. Gupta, A. F. Dana, J. P. Hespanha, R. M. Murray, and B. Hassibi,\ ``Data transmission over networks for estimation and control,"
\emph{IEEE Trans. Automat. Contr.}, \relax vol. 54, no. 8, pp. 1807-1819, Aug. 2009.

\bibitem{IRef24}
L. Schenato, B. Sinopoli, M. Franceschetti, K. Poolla, and S. S. Sastry,\ ``Foundations of control and estimation over lossy networks,"
\emph{Proc. IEEE}, \relax vol. 95, no. 1, pp. 163-187, Jan. 2007.

\bibitem{IRef25}
K. Xin, X. Cao, J. Chen, P. Cheng, and L. Xie,\ ``Optimal controller location in wireless networked control systems,"
\emph{Int. J. Robust Nonlinear Control}, \relax pp. 1-19, Oct. 2013.

\bibitem{IRef26}
H. Gao, T. Chen, and J. Lam,\ ``A new delay system approach to network-based control,"
\emph{Automatica}, \relax vol. 44, no. 1, pp. 39-52, Jan. 2008.

\bibitem{IRef27}
H. Hur and H. S. Ahn,\ ``Discrete-time $H_\infty$ filtering for mobile robot localization using wireless sensor network," \emph{IEEE Sensors J.},
\relax vol. 13, no. 1, pp. 245-22, Jan. 2013.

\bibitem{IRef28}
U. Tiberi, C. Fischione, K. H. Johansson, and M. D. Di Benedetto,\ ``Energy-efficient sampling of networked control systems over IEEE 802.15.4 wireless networks," \emph{Automatica},
\relax vol. 49, no. 3, pp. 712-724, Mar. 2013.

\bibitem{IRef29}
R. M. Murray,\ ``Recent research in cooperative control of multivehicle systems,"
\emph{ASME Journal of Dynamic Systems, Measurement, and Control}, \relax vol. 129, no. 5, pp. 571-583, May 2007.

\bibitem{IRef30}
J. C. Engwerda, \emph{Linear Quadratic Dynamic Optimization and Differential Game Theory}, \relax Chichester: Wiley, 2005.

\bibitem{IRef31}
E. Semsar-Kazerooni and K. Khorasani,\ ``Multi-agent team cooperation: A game theory approach,"
\emph{Automatica}, \relax vol. 45, no. 10, pp. 2205-2213, Oct. 2009.

\bibitem{IRef32}
H. Mukaidani,\ ``Local feedback Pareto strategy for weakly coupled large-scale discrete-time stochastic systems,"
\emph{IET Contr. Theory Appl.}, \relax vol. 5, no. 17, pp. 2005-2014, Nov. 2011.

\bibitem{IRef33}
R. Olfati-Saber and R. M. Murray,\ ``Consensus problems in networks of agents with switching topology and time-delays,"
\emph{IEEE Trans. Automat. Contr.}, \relax vol. 49, no. 9, pp. 1520-1533, Sep. 2004.

\bibitem{IRef34}
A. Ulusoy, O. Gurbuz, and A. Onat,\ ``Wireless model-based predictive networked control system over cooperative wireless network,"
\emph{IEEE Trans. Ind. Inf.}, \relax vol. 7, no. 1, pp. 41-51, Feb. 2011.

\bibitem{IRef35}
X. Liu and A. Goldsmith,\ ``Wireless medium access control in networked control systems,"
\emph{in Proc. IEEE Amer. Contr. Conf.}, \relax pp. 688-694, 2004.

\bibitem{IRef36}
M. Pajic, S. Sundaram, G. J. Pappas, and R. Mangharam,\ ``The wireless control network: A new approach for control over networks,"
\emph{IEEE Trans. Automat. Contr.}, \relax vol. 56, no. 10, pp. 2305-2318, Oct. 2011.

\bibitem{IRef37}
Z. Wang, D. Ding, H. Dong, and H. Shu,\ ``$H_\infty$ consensus control for multi-agent systems with missing measurements: The finite-horizon case,"
\emph{Syst. Contr. Lett.}, \relax vol. 62, no. 10, pp. 827-836, Oct. 2013.

\bibitem{IRef38}
Y.-P. Tian and C.-L. Liu,\ ``Consensus of multi-agent systems with diverse input and communication delays,"
\emph{IEEE Trans. Automat. Contr.}, \relax vol. 53, no. 9, pp. 2122-2128, Oct. 2008.

\bibitem{IRef39}
V. S. Vokharaie, O. Mason, and M. Verwoerd,\ ``D-stability and delay-independent stability of homogeneous cooperative systems,"
\emph{IEEE Trans. Automat. Contr.}, \relax vol. 55, no. 12, pp. 2882-2885, Dec. 2010.

\bibitem{IRef40}
H. Li, L. Lai, and H. V. Poor,\ ``Multicast routing for decentralized control of cyber physical systems with an application in smart grid,"
\emph{IEEE J. Sel. Areas Commun.}, \relax vol. 30, no. 6, pp. 1097-1107, Jul. 2012.



\bibitem{IRef43}
K.~J.~Astrom, \emph{Introduction to Stochastic Control Theory},
\relax New York: Academic Press, 1970.

\bibitem{IRef44}
K. J. Astrom and B. Wittenmark, \emph{Computer-Controlled Systems Theory and Design},
\relax Prentice Hall, 3rd edition, 1997.

\bibitem{IRef45}
L. Dong and Y. Zhang,\ ``On design of a robust load frequency controller for interconnected power systems,"
\emph{in Proc. IEEE Amer. Contr Conf.}, \relax pp. 1731-1736, 2010.

\bibitem{IRef46}
C. E. Fosha and O. I. Elgerd,\ ``The megawatt frequency control problem: A new approach via optimal control theory,"
\emph{IEEE Trans. Power App., Syst.}, \relax vol. PAS-89, no. 4, pp. 563-577, Apr. 1970.

\end{thebibliography}
\end{document}